\documentclass[showpacs,amsmath,amssymb,preprint]{revtex4}
\usepackage{amsmath}    % need for subequations
\usepackage{graphicx}   % need for figures
\usepackage{epsfig}   % need for figures
\usepackage{verbatim}   % useful for program listings
\usepackage{color}      % use if color is used in text
\usepackage{subfigure}  % use for side-by-side figures

\begin{document}

\title{Transition of a particle between adjacent optical traps: A study using catastrophe theory}
\author{Deepak Kumar,  Shankar Ghosh  and S. Bhattacharya}
\affiliation{Department of Condensed Matter Physics and Materials Science,~ Tata Institute of Fundamental Research,~ Homi Bhabha Road,~ Mumbai 400-005,~ India\\}

%\pacs{46.55.+d}{}
%\pacs{nn.mm.xx}{Second pacs description}
%\pacs{nn.mm.xx}{Third pacs description}

\begin{abstract}
In spite of the widespread use of optical tweezers as a quantitative tool to measure small forces, there exists no unambiguous and simple experimental method for either validating its theoretically predicted form or empirically parameterizing it over the entire range. This problem is addressed by studying the transition of a colloidal particle between two spatially separated optical traps. The transition as a function of the relative intensity of the traps and the separation between them reveals a formal resemblance to the `butterfly catastrophe'  which also maps onto to phase transitions observed, for example in ferroelectrics, on a phenomenological level. The method has been used to experimentally determine the force-displacement curve for an optical trap over its entire range.

\end{abstract}
\maketitle

\maketitle

The optical tweezer is widely regarded as an important experimental tool in the field of soft matter and biological sciences. It is frequently used to noninvasively measure small forces ($\sim  10^{-12}$ N) and to transport matter with precision \cite{Grier:tweezer, Dholakia:tweezer,Dholakia:beam:shaping}. The trapping potential due to the tweezer is usually assumed to be harmonic. Although this approximation holds true for small displacements from the center of the trap, significant deviations are expected for large displacements\cite{dunlop}. A detailed form of the trapping potential can be derived using scattering theory \cite{nieminen_optical_2007}. However, only a few reports compare the experimentally determined form of the entire potential of a single trap with the calculated one \cite{jahnel_measuring_2011} and none compare the scenario when the trapping potential is formed by the superposition of two or more traps. A direct comparison is experimentally difficult because it requires enabling the trapped particle to explore the energetically unfavorable high lying states of a single well in a repeatably controlled manner.

We circumvent the difficulties by studying the transition of a colloidal particle between \emph {two} spatially separated optical traps, created using an spatial light modulator (SLM), as a function of the relative intensity  ($p$) of the two traps. For small separations, the passage of the particle from one trap to the other is continuous. As the separation is increased the transition becomes discrete consisting of single or multiple jumps. In this regime the position of the particle is relatively insensitive to changes in $p$ except near the point of jump, where for a small incremental change in $p$, a large change in the position occurs. Such sudden changes caused by smooth alterations in the control parameter are observed in a variety of natural phenomenon, e.g., Euler buckling of a beam or electronic flip-flop circuit \cite{pippard_response}, and are categorized under the common name "catastrophe" \cite{zeeman_catastrophe_1977,pippard_response,gilmore_catastrophe_1993}. 
The measured equilibrium position of the particle as a function of the relative intensity of the two traps and the separation between them has been used to calculate the force-displacement curve for an optical tweezer over a very large spatial extent. Additionally, the experimental protocol provides, to our knowledge, a novel method to translate or position a particle with sub-pixel resolution of the SLM.

\begin{figure} [p]
\includegraphics [width=14cm]{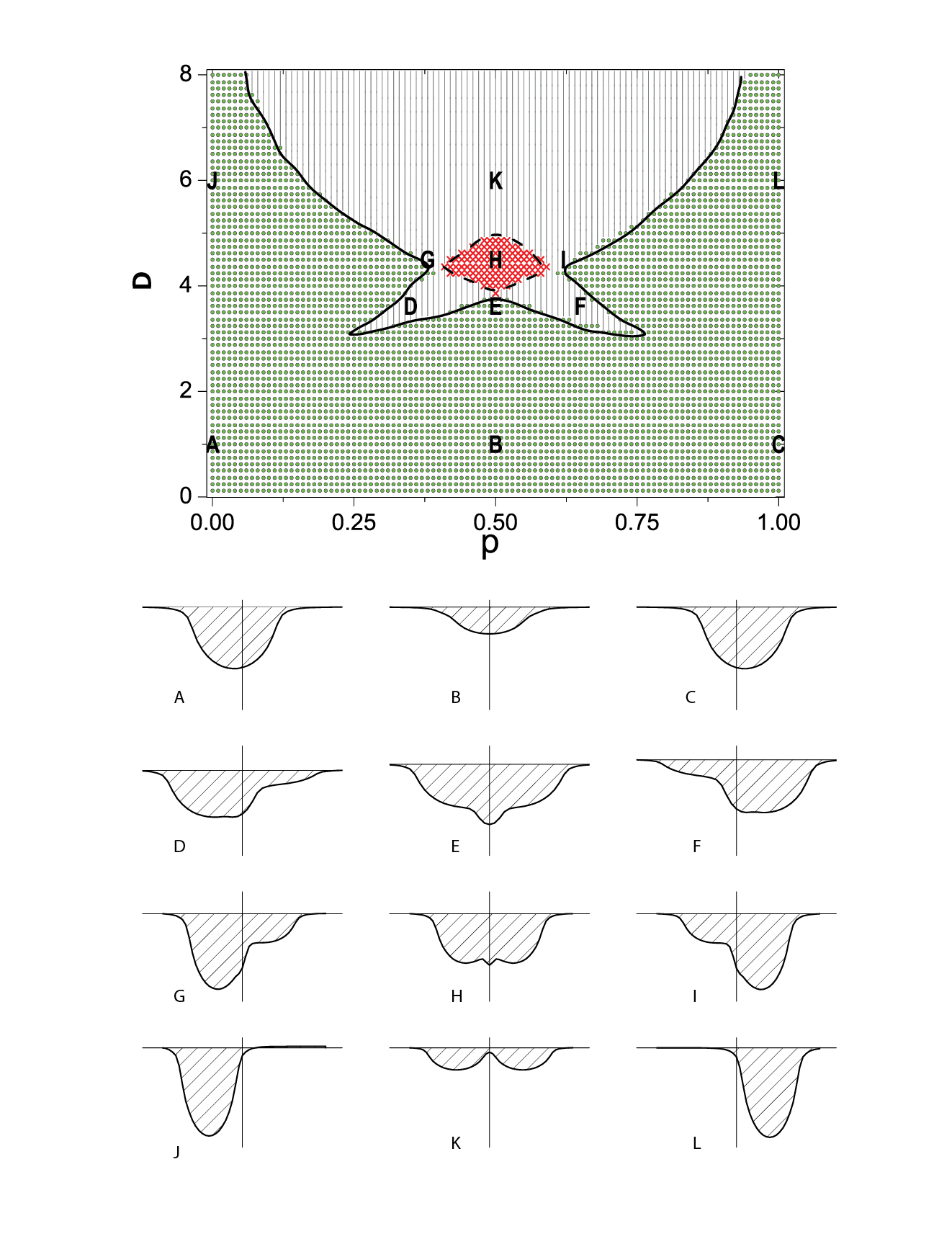}
\caption{ (Color online) Phase diagram showing the number of minima in the potential for various values of the relative intensity, p, and separation, D; circles correspond to one minimum, vertical lines to two minima and crosses to more than two minima respectively. The form of potential at different points in the $p-D$ plane marked by $A \cdots  L$ are shown in the bottom panels. In the above figure $D$ is plotted in units of $\lambda$.}
\label{phase diagram}
\end{figure}

The working principle of the method relies centrally on the fact that an optical tweezer is not purely harmonic for large displacements from equilibrium, as mentioned above and that a combination of two spatially separated optical traps would yield an effective potential whose functional form in space can be varied\cite{dunlop}.
By changing the relative intensity ($p$) of two traps placed at distance $D$ (measured in units of $\lambda$ ($=1.064 \mu m$), the wavelength of the trapping laser),  various forms of effective potentials can be generated. We estimate the form of these potentials from scattering theory using the Matlab based ``optical tweezers computational toolbox''  developed by Nieminen  \textit{et. al.} \cite {nieminen_optical_2007}. For particle radius, $R>\lambda$ $R=2.5 \mu m$ in our case) two clear trends are observed. For $D$ smaller than a critical distance $D_c \sim 3 \lambda$, i.e., $D<D_c$, the number of minima in the potential, n=1. In this regime as $p$ is changed from 0 to 1 the  minimum moves continuously from the center of one well to that of the other, \emph{reversibly} translating the particle with it. For $D>D_c$  the particle continues to occupy its initial energy state until it becomes a stationary point of inflection in space. The particle then makes an abrupt and hysteretic transition to the the next available potential energy minimum. The corresponding phase diagram as a function of the control parameters $D$ and $p$ is shown in the top panel of Fig. \ref{phase diagram}. The circles, vertical lines and crosses, correspond to situations where $n=1$, $n=2$ and $n>2$, respectively.
The boundary line which separates the region $n=1$ from the region $n>1$ is marked by a solid line in Fig. \ref{phase diagram}. 
The dashed line in the figure separates the parameter space corresponding to $n>2$ from $n=2$.
The lower panels of Fig.\ref{phase diagram} show the form of the potential for different values of the control parameters marked in the phase diagram by alphabetic letters ranging from A to L. We note that if $R<\lambda$ the phase diagram has a cusp separating the region of one minimum from the region of two minima; however, in this paper we restrict ourselves to the $R>\lambda$-case.

The phase diagram obtained above suggests that the phenomena can be described in terms of a "butterfly catastrophe" \cite{zeeman_catastrophe_1977,castellano_catastrophe_2007,gilmore_catastrophe_1993,kitano_optical_1981}, one of the seven elementary catastrophes. It can be derived from a six degree polynomial $U(x; a,b,c,d)=\frac{1}{6}x^6-ax-\frac{1}{2}bx^2-\frac{1}{3}cx^3-\frac{1}{4}dx^4$ where a, b, c, d are numerical coefficients. The separatrix of the catastrophe defined by $\frac{\partial U}{\partial x}= 0$ and $\frac{\partial^2 U}{\partial x^2}= 0$ separates  the control parameter space  into regions where $U$ has  one, two and three minima, respectively. By comparing our phase diagram with that obtained for the butterfly catastrophe\cite{zeeman_catastrophe_1977}, we can make the following correspondence between the parameters in our experiment and the coefficients of the polynomial given above:$x\sim$ position of the particle along the line joining the two traps, $p\sim a$, $D\sim b$, $c=0$ in our case. 

It is also interesting to note that the phenomenological "Landau-Devonshire" model with generalized Landau approach that accounts for both first order and second order phase transitions, assumes a similar form for the free energy as a function of the order parameter, with the coefficient $c=0$ for symmetry reasons \cite{Devonshire, _physics_2007}. This leads us to an analogy between the present experiment and the more familiar and physical case of a phase transition. The position of the particle in our experiment is like an order parameter in a phase transition, $p$ is like a bias factor or an external field, the distance between the two traps is like the temperature. $p=0.5$ corresponds to the situation where there is no external force. Moving along the $p=0.5$ line for $R>\lambda$, the situation considered in this paper, $x(D)$ shows a first order transition. However, for $R<\lambda$ the transition becomes second order\cite{dunlop} and the phase diagram shown in Fig. \ref{phase diagram} (top panel) reduces to a cusp. In $U(x;c=0)$ above the transition is first order for $d>0$ and second order for $d<0$. 
The functional form $U(x;c=0)$ describes the phase diagram well, however it holds true only for small values of the order parameter that is only near the point of transition.

\begin{figure} [p]
\includegraphics [width=14cm]{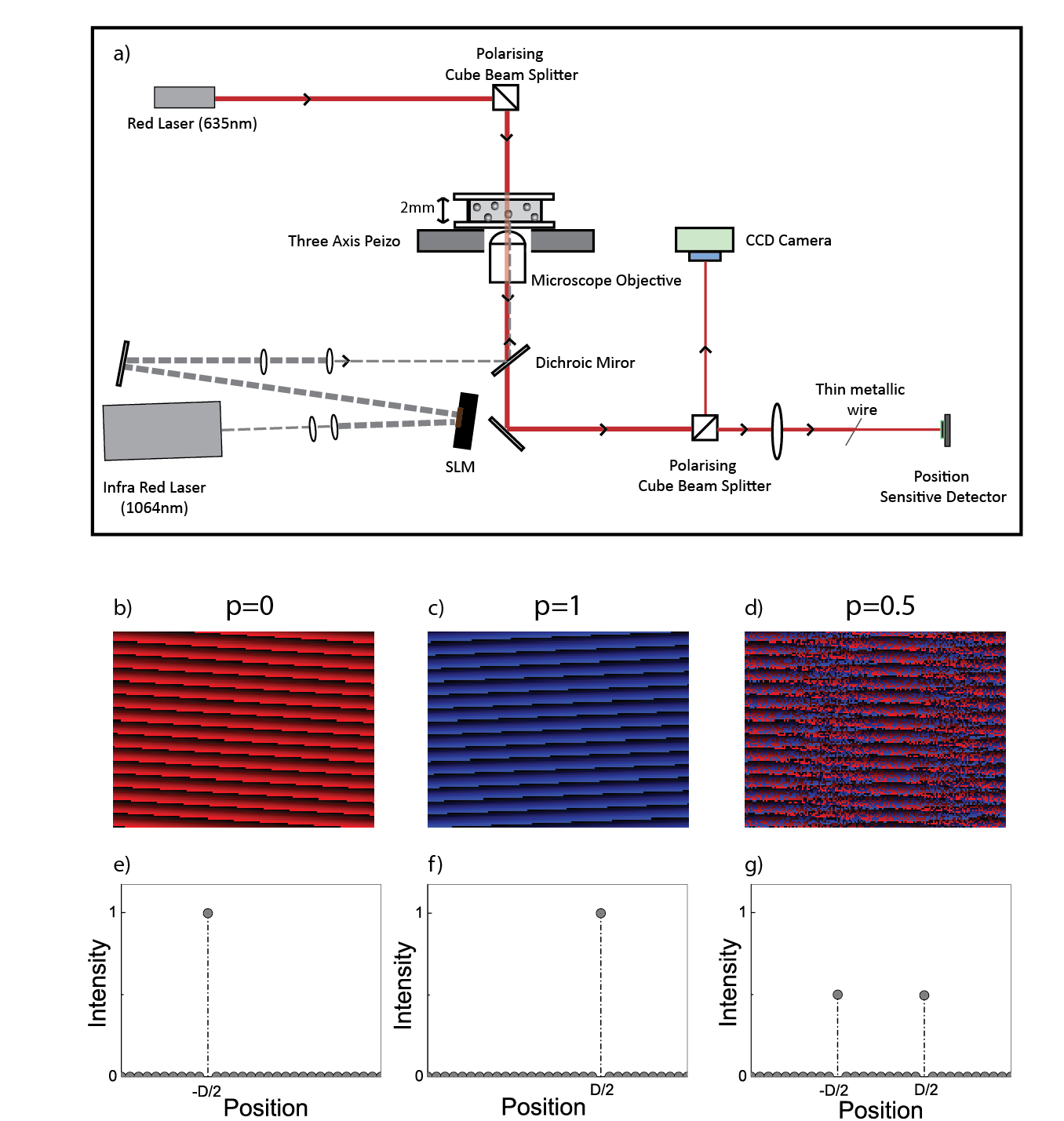}

\caption{(Color online) a) The experimental setup. A 1064 nm laser beam is used to form optical traps. The incident wavefront is phase-modulated using an SLM to form multiple traps. The sample consists of 5 micron Silica microspheres in an aqueous environment. b) and c) show holograms generated using the Gerchberg-Saxton algorithm to create single optical traps. d) is the hologram obtained by combining b) and c) at a pixel fraction, $p=0.5$, using the random masking algorithm. The holograms are gray-scale images, (false color in the online version has been used for clarity). The corresponding bottom panels e), f) and g) respectively show the simulated intensity profile obtained by taking the Fourier transform of unit amplitude complex numbers corresponding to the phase in the holograms.}
\label{setup}
\end{figure}

The experimental setup is schematically shown in Fig.\ref{setup} (a). A Nd:YAG infrared laser of wavelength, $\lambda$=1064nm focused using a 12 NA, 63x, water immersion objective (Carl Zeiss Objective "C-Apochromat"), is used to generate the optical trap. A phase only spatial light modulator (Hamamatsu LCOS-SLM) generates multiple optical traps by modulating the phase of the incident wavefront. The sample consists of 5 micrometer (diameter) silica (refractive index =1.5) particles suspended in an aqueous solution sealed in a sample cell made of clean microscope cover slips, for the bottom and the top plates and a rubber O-ring for the side walls. We use the Gerchberg-Saxton algorithm (G-S algorithm) to generate the phase holograms corresponding to a single optical trap \cite{GS, Grier:SLM}. The algorithm consists of an iterative procedure that converts an intensity-image into its corresponding phase-only hologram\cite{GS}. We generate two phase holograms, shown in Figure 2b and Figure 2c, corresponding to two optical traps separated spatially by a distance $D$. These two holograms are then combined  using a modification of the random masking algorithm proposed by Montes et. al. \cite{montes-usategui_fast_2006}. This method is based on the idea that even a part of a hologram generates the whole image back and hence one can combine two holograms by taking half of each and merging them to form a single hologram that recreates both the images. In this method we randomly choose a fraction of pixels $p$ from the total available pixels on SLM, and assign to these pixels, values corresponding to the first hologram and the remaining $1-p$ fraction of pixels to values corresponding to the second. The fraction p determines the relative intensity of traps. The intensity varies linearly with p; for $p=0.5$ the intensities of the traps are equal. The corresponding phase hologram is shown in Fig.\ref{setup} (d). We can simulate the intensity profile obtained in the image plane from a hologram $\phi (x,y)$ by taking the Fourier transform of $e^{i\phi (x,y)}$. In the panels (e), (f) and (g) of Fig.\ref{setup}, we show the intensity profile calculated in this way for holograms (b), (c) and (d) respectively. The traps are placed at $-D/2$ and $D/2$.

\begin{figure} [p]
\includegraphics [width=14cm]{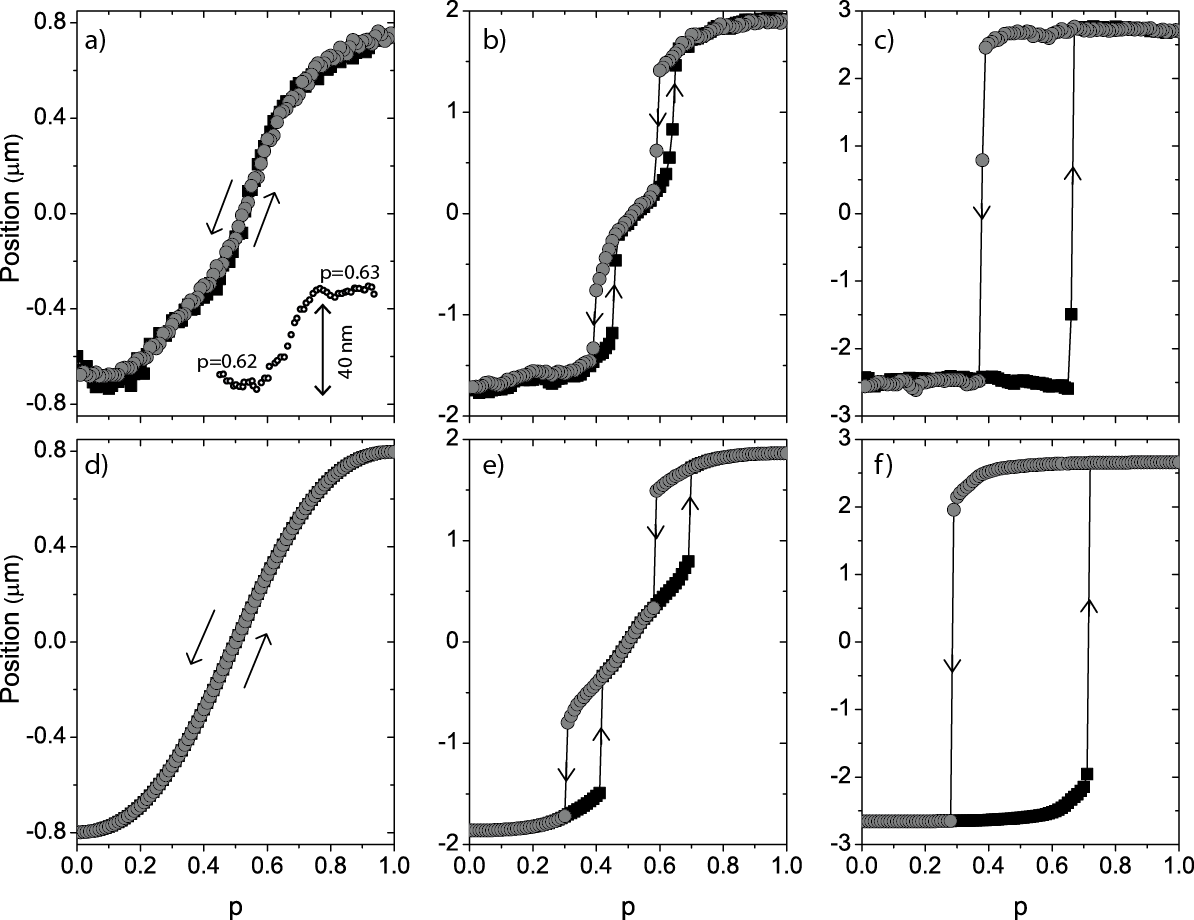}
\caption{Motion of the particle as the relative intensity between the two traps is changed, for separations $D\lambda$= 1.5$\mu m$, 3.6 $\mu m$ and 5 $\mu m$ is shown in a), b) and c) respectively. The onward run is shown in square (black) while the return run in circles (grey). The corresponding bottom panels are the curves obtained from numerical calculations for identical parameters. The inset to panel a) shows the position as a function of time as p is changed from 0.62 to 0.63.  The corresponding change in position is $40nm$.}
\label{data}
\end{figure}

Figure\ref{data} (a)  shows the position of a trapped particle along the direction in which the two traps have been placed, as a function of  increasing (squares) and decreasing(circles) values of $p$ (bias factor). The panels (a), (b) and (c) correspond to  $D\lambda = 1.5$, 3.6, and $5 \mu m $  respectively. The directions of increasing and decreasing $p$ are shown by the analogously pointing arrows. For $D \lambda=1.5 \mu m < D_c\lambda$, the particle moves \emph{continuously and reversibly} from the center of one trap to that of the other as $p$ is changed from 0 to 1.  The non-linear variation in the position as a function of $p$ is a manifestation of the anharmonicity of the trap. The inset to Fig.\ref{data} (a)  shows that the  method can  position an optically trapped particle spatially with a precision of $40 nm$, significantly better than a precision of $250 nm$ achievable by shifting the trap by one pixel on the SLM. This clearly demonstrates the usefulness of this method to achieving particle positioning with great precision.

For $D>D_c$ the transition of the particle from the initial to final state happens in a discontinuous and hysteretic manner, i.e, the external field-induced transition is first-order. Fig. \ref{data} (b) and (c) show the transition for\ $D\lambda=3.6\mu m$ and $D\lambda=5\mu m$ respectively. While in the later case the transition happens in a single step, the former shows two steps. The two-step transition arises because in this range of parameter values there exists a third minimum in the potential about $x=0$ , a feature of the 'butterfly catastrophe'  for c=0 and $d>0$\cite{gilmore_catastrophe_1993}. Interestingly, $BaTiO_3$, a ferroelectric which undergoes first-order phase transition modeled by the Landau-Devonshire theory, shows a similar double hysteresis loop in the polarization-electric field curve at a temperature slightly greater than the critical temperature, $T_c$  \cite{Double_Hysteresis}.
Fig.\ref{data} (d), (e) and (f) show the position of the particle as a function of $p$ for $D\lambda=1.5$, $3.6$ and $5 \mu m$, respectively, computed numerically using the algorithm given by  Nieminen  \textit{et. al.} \cite {nieminen_optical_2007}. The agreement with the experimental findings is high (see Fig.\ref{data} (a), (b), (c)).The extent to which the trapping potential is captured by the scattering theory is reflected in the matching of the numerical calculation with the experimental data. Thus the variation of the position with $p$ can be used to validate the accuracy of a calculated trapping potential. Further it also allows one to parametrize an empirical form of it. 

\begin{figure} [p]
\includegraphics [width=14cm]{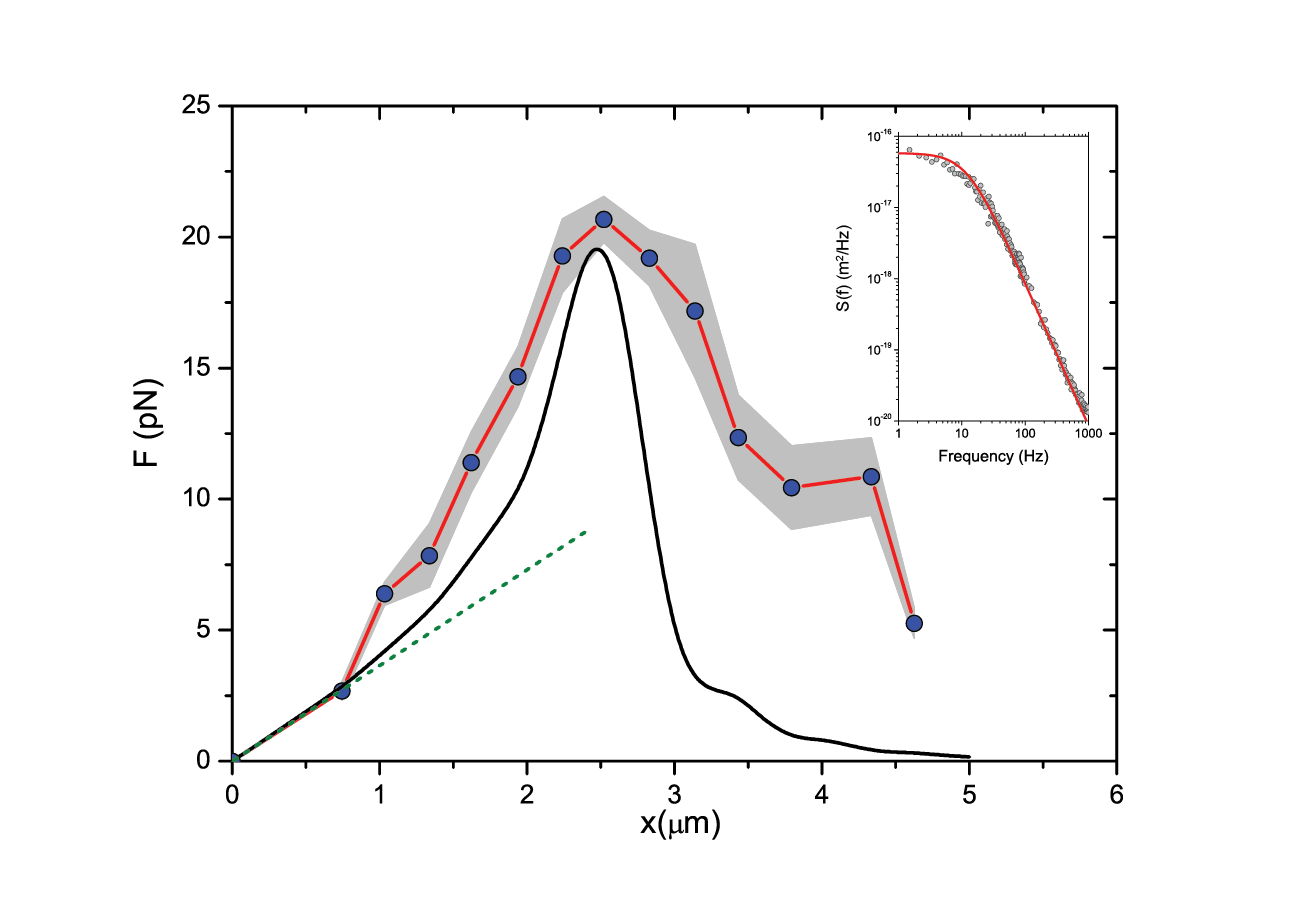}
\caption{(Color online) The force-displacement curve for an optical tweezer obtained from $x$ vs. $p$ curves shown in Fig.\ref{data} using the method described in the text, is shown by circles connected with straight line segments. The scatter plot in the inset shows the power spectrum of fluctuation in position in the configuration p=0. The solid line is a lorentzian fit to the data giving a corner frequency of 12 Hz. Using this we calculate the small displacement limit (dotted line) of optical trap stiffness which has been used to provide the physical scale to the experimentally obtained force displacement curve. The curve obtained from numerical calculations employing scattering theory is shown as solid line. }
\label{FD}
\end{figure}

Alternatively, one can determine the exact form of the trapping potential from the experimental data using the following method. From the various runs like those shown in Fig.\ref{data} (a), (b) and (c), we obtain the equilibrium position of the particle as a function of the parameters $p$ and $D$. We assume that the force-displacement curves for the two traps have an identical form $F(x)$. The net force acting on a particle in the presence of the two traps is, then, of the form: $(1-p) F(x)+pF(x-D)$. For every equilibrium position $x(p,D)$ of the particle measured at a given $p$ and $D$, the net force acting on it due to the two optical traps add up to zero, i.e., $(1-p)F(x)+pF(x-D)=0$.  Discretizing the full range of $x$ accessible in our experiment, we obtain a set of linear homogeneous equations. We further assume that along any given direction the function $F(x)$ is odd about the center of the trap, i.e., $F(-x)=-F(x)$, i.e., the potential is invariant under an inversion operation through the center of the optical trap and solve the equations to determine $F(x)$ up to an unknown overall multiplication factor. To fix this multiplication factor we have used the power spectrum of fluctuation in position of a trapped particle in the configuration $p=0$ shown as scatter plot in the inset to Fig.\ref{FD}. A lorentzian fit to the power spectrum (solid line),  $S(f)=\frac{k_BT}{\Gamma \pi^2(f^2+f_c^2)}$, where $\Gamma=6\pi\eta R$, $\eta=1mPas$ is the viscosity of the surrounding medium,  gives a corner frequency, $f_c=12Hz$. The trap stiffness in the small displacement limit is obtained from the corner frequency using the relation $k_{opt}=12 \pi ^2 \eta R f_c$ \cite{neuman:2787}. The value of this trap stiffness has been used to fix the slope of the force-displacement curve in the small displacement limit (shown as a dotted line).The resulting force-displacement curve in physical units (circles connected by straight line segments) is shown in Fig.\ref{FD}; the curve obtained from numerical calculations employing scattering theory using the algorithm developed by Nieminen \textit{et. al.} \cite {nieminen_optical_2007} is shown as solid line.

\begin{figure} [p]
\includegraphics [width=14cm]{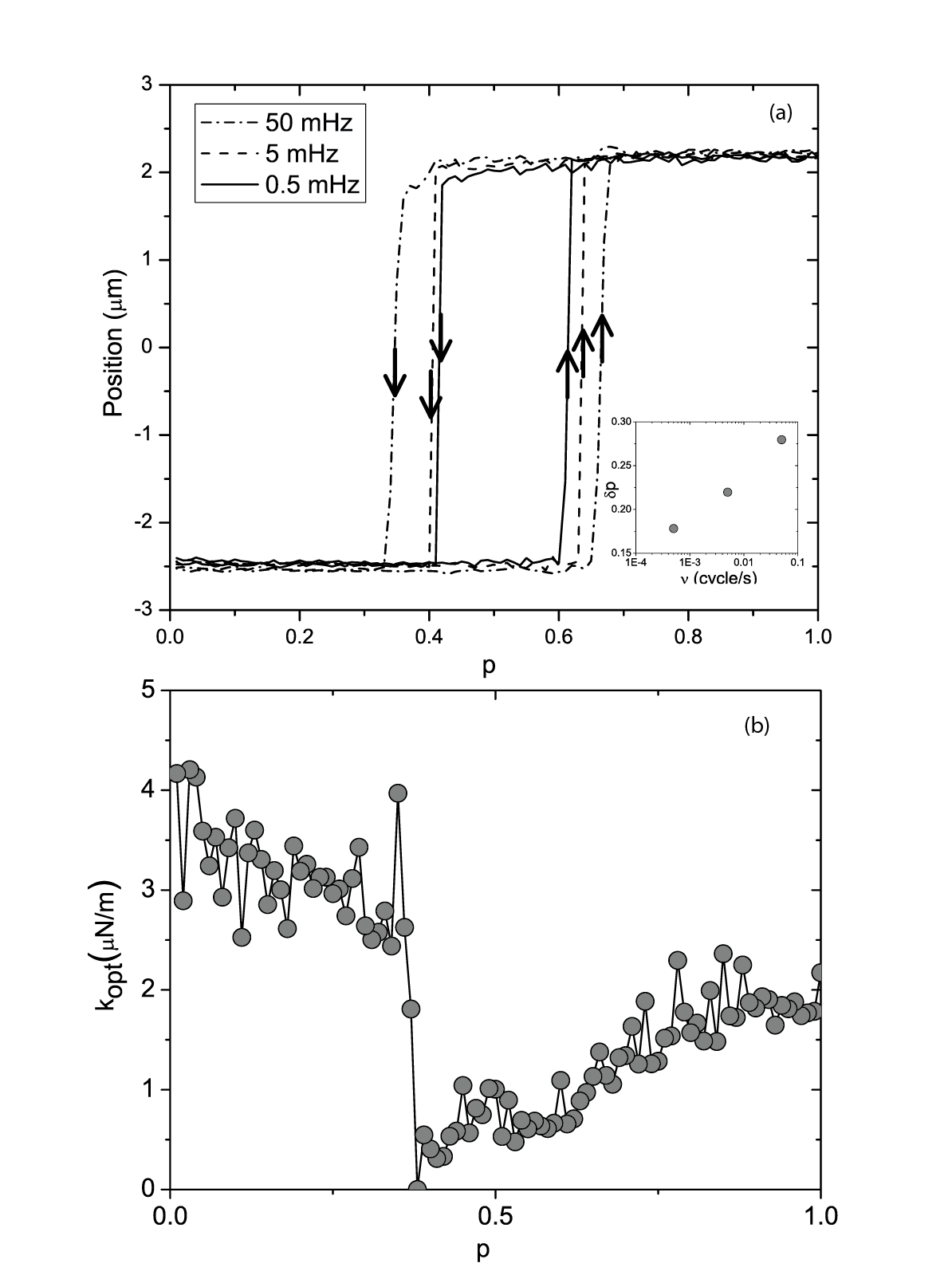}
\caption{(a) Hysteresis loops done at three different time-rates. The solid, dash, dash-dot lines correspond to 0.5 mHz, 5 mHz and 50 mHz respectively. The width of the hysteresis loop, $\delta p$, increases with increasing rate as shown in the inset implying a finite role of temperature. (b) shows the trap stiffness, $k_{opt}$, obtained from thermal fluctuations in position as a function of p for $D\lambda$= $5 \mu m$ respectively. Only the return part of the cycle is shown.}
\label{hysteresis}
\end{figure}

According to the delay convention \cite{zeeman_catastrophe_1977, gilmore_catastrophe_1993}, a particle continues to occupy a potential minima until it becomes a point of inflection as mentioned above. However, in a system with finite thermal effects, the transition may happen earlier owing to fluctuations and in such a situation the width of the hysteresis loop, $\delta p$ would depend on the sweep rate of $p$ \cite{bertotti_hysteresis_1998}. In Fig.\ref{hysteresis} (a), we show the hysteresis loops for $D\lambda = 4.7 \mu m$ 
done at three different rates, $\nu$ =50 mHz, 5mHz and 0.5 mHz. The width of the hysteresis loop is observed to increase with increasing sweep rate as shown in the inset. The log-linear relationship between the rate and the width of the loop implies an activated response, as expected \cite{bertotti_hysteresis_1998}.

A system makes a transition from one state to another when the initial state becomes unstable, i.e., the stiffness  ($k_{opt}$) associated with the potential, defined as the second  derivative of the potential with respect to position calculated at this minima, goes to zero. Fig.\ref{hysteresis} (b) shows $k_{opt}$ calculated in our experiment from the temporal fluctuations in the position using, $k_{opt} = k_B T/\sqrt{<\delta x^2>}$ \cite{sharma_microrheology_2008}. 

\begin{figure} [p]
\includegraphics [width=20cm]{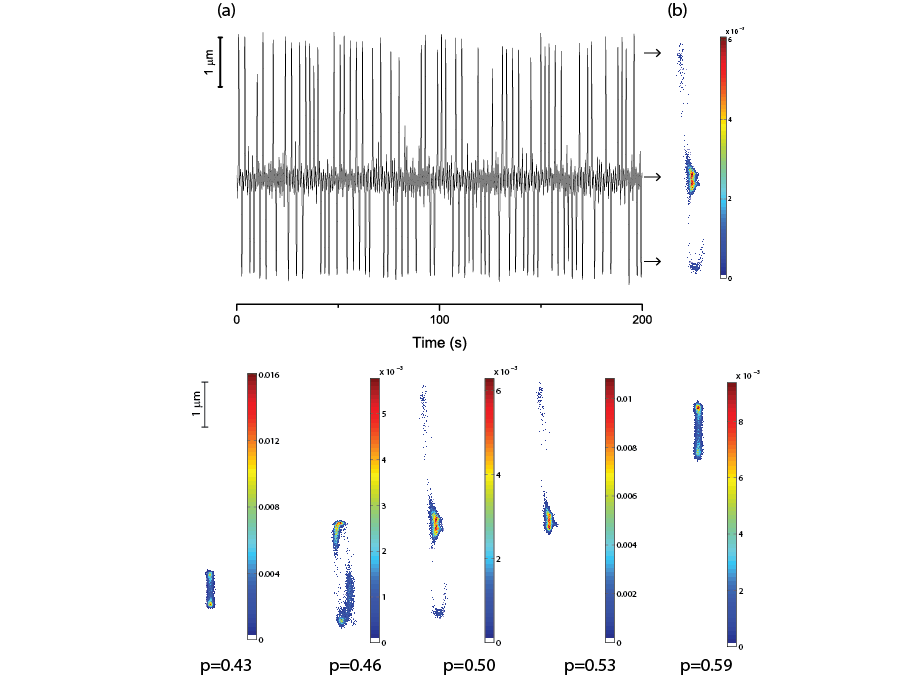}
\caption{(Color online) (a) time series of the position in presence of an oscillatory perturbation at 1 Hz in the configuration $D\lambda=4\mu m$ and $p=0.50$. (b) the spatial distribution of the residence time obtained from the data shown in (a) normalized with the total time of measurement. The bottom panel shows similarly obtained distributions of the residence time, for $p=0.43, 0.46, 0.50, 0.53, 0.59$ respectively. The dumbbell shaped distribution in any one potential minimum is a consequence of the fact that in a sinusoidal oscillation a particle spends more time towards the extreme than in the center.}
\label{noise}
\end{figure}

For the parameter space marked as "H" in Fig.\ref{phase diagram} one expects to observe simultaneous occurrence of three minima. To explore these minima we have applied an additional sinusoidal perturbation to the system. The sample cell containing the trapped particle was oscillated using a peizo-stage at a frequency of 1 Hz and amplitude $2 \mu m$. The time trace of position of such a system in the configuration $D\lambda=4 \mu m$ ($D=3.76$) and $p=0.50$ is shown in Fig.\ref{noise} (a). We observe that the particle is in the central minimum most of the time but occasionally toggles between the upper and the lower well in an approximately stochastic manner implying the importance of thermal fluctuations. Fig.\ref{noise} (b) shows the residence time distribution obtained from \ref{noise} (a) and normalized with respect to the total time of measurement. The distribution is clustered at three positions implying three minima in the potential. The bottom panel of Fig.\ref{noise} shows the evolution of similarly obtained residence time distribution as $p$ is changed through the values 0.43, 0.46, 0.50, 0.53 and 0.59 respectively. The dumbbell shaped distribution obtained in any one potential minimum is a consequence of the fact that in a sinusoidal oscillation a particle spends more time towards the extreme than in the center.

In conclusion, we have shown that the behavior of a single particle trapped in a potential well formed by combining two spatially separated optical traps as a function of their relative intensity and separation resembles a butterfly catastrophe formalism\cite{gilmore_catastrophe_1993} and is also similar to the physical phenomenon of a phase transition as described in the Landau-Devonshire theory \cite{Devonshire,_physics_2007}. The measured equilibrium position of the particle as a function of the control parameters has been used to calculate the force-displacement curve over a large spatial extent, not easily accessible in single trap experiments. It provides a much needed benchmark with which the calculated potential in complex optical trapping scenarios \cite{Dholakia:beam:shaping} can be compared. Furthermore, the technique of combining two optical traps reported in this paper can be used to create various shapes of the trapping potential and provides a method to translate the trapped particle with sub-pixel resolution of the SLM. These results are clearly useful in simultaneously manipulating a multi-particle system with high precision.

We thank Prof. Rajaram Nityananda for useful discussions.

\end{document}